\documentclass[a4paper,11pt]{article}

\usepackage{jinstpub} 

\title{The impact of improved plasma diagnostics on modeling the X-ray Universe}

\author[a,b,1]{Junjie Mao,\note{Corresponding author.}}
\author[c,d,b]{Fran\c{c}ois Mernier,}
\author[b,e]{Jelle S. Kaastra,}
\author[f,b]{Liyi Gu,}
\author[b]{Missagh Mehdipour,}
\author[b]{and Jelle de Plaa}

\affiliation[a]{Department of Physics, University of Strathclyde, Glasgow G4 0NG, UK}
\affiliation[b]{SRON Netherlands Institute for Space Research, Sorbonnelaan 2, 3584 CA Utrecht, the Netherlands}
\affiliation[c]{MTA-E\"{o}tv\"{o}s University Lend\"{u}let Hot Universe Research Group, P\'{a}zm\'{a}ny P\'{e}ter s\'{e}t\'{a}ny 1/A, Budapest, 1117, Hungary}
\affiliation[d]{Institute of Physics, MTA-E\"{o}tv\"{o}sy,P\'{a}zm\'{a}ny P\'{e}ter s\'{e}t\'{a}ny 1/A, Budapest, 1117, Hungary}
\affiliation[e]{Leiden Observatory, Leiden University, PO Box 9513, 2300 RA Leiden, the Netherlands}
\affiliation[f]{RIKEN High Energy Astrophysics Laboratory, 2-1 Hirosawa, Wako, Saitama 351-0198, Japan}

\emailAdd{junjie.mao@strath.ac.uk}

\abstract{The high-resolution X-ray spectrum of the Perseus galaxy cluster observed with the \textit{Hitomi} satellite challenges astrophysical collisional ionized plasma models that are widely used in the community.  Although \textit{Hitomi} spun out of control, several \textit{Hitomi}-level missions have been proposed and some funded. The spectrometers aboard these future missions have a broader energy range and/or a higher spectral resolution to achieve different scientific goals. Accurate plasma models and atomic data are crucial for plasma diagnostics of high-quality spectra. Here, we present a few cases where improvement of plasma diagnostics will be decisive to better understand celestial bodies and their physical processes at play. We focus on collisional ionized and photoionized astrophysical plasmas in the context of developments of plasma models, as well as current and future generations of spectrometers.}

\keywords{Plasma diagnostics - probes; Data analysis; Spectrometers; X-ray detectors and telescopes}

\proceeding{3$^{\text{rd}}$ European Conference on Plasma Diagnostics (ECPD 2019)\\
  From May 6$^{\text{th}}$ to May 9$^{\text{th}}$ 2019\\
  Lisbon, Portugal}

\begin{document}
\maketitle
\flushbottom

\section{Hot plasmas in the Universe}
\label{sec:intro}
Hot plasmas can be found in a wide range of astrophysical environments. For instance, stellar coronae, the accretion flow of the supermassive black holes at the center of nearly all galaxies, the intracluster media among member galaxies within galaxy clusters, and the warm-hot intergalactic media (i.e., the cosmic web filaments) among clusters of galaxies. In fact, about half of the baryons in the entire Universe are expected to be in the form of hot plasmas with a temperature of millions to billions of degree and shine in the X-ray wavelength range (0.1--100~\AA). That is to say, X-ray observational facilities hold the key to probe the nature of these hot plasmas.

Spectroscopy, as one of the main techniques to study the electromagnetic radiation of the Universe, is used to constrain physical properties (temperature, density, chemical composition, microscopic turbulence, line of sight velocity, etc.) of the observing targets. Plasma diagnostics in the X-ray band provide important information to answer major astrophysical questions involving hot plasmas like ``How does AGN feedback influence the host galaxies?", ``What is the chemical composition of the Universe and what is the origin of the observed elements?", and ``Where are the missing baryons in the local Universe?". 

In 1975, the very first X-ray spectrum ($1.3-16$~keV) of the Perseus galaxy cluster was obtained with the collimated proportional counter aboard \textit{Ariel 5}. As shown in \cite{mit76}, a broad emission feature peaking around $\sim$7~keV stands out above the continuum. This emission feature was interpreted as (unresolved) transitions from H-like and He-like Fe ions, while the continuum was described as the thermal bremsstrahlung emission of the hot intracluster medium. About four decades later, the Perseus galaxy cluster was revisited by \textit{Hitomi} (previously known as ASTRO-H \cite{tak14}). The Soft X-ray Spectrometer (SXS) aboard \textit{Hitomi} achieves an unprecedented energy resolution of 5~eV throughout its soft X-ray band. As a result, the ``single" emission feature observed with \textit{Ariel 5} in the $5-8$~keV is resolved as multiple emission lines: a prominent He$\alpha$ triplet from Fe XXV ($\sim6.5-6.6$~keV), the Ly$\alpha$ line from Fe XXVI ($\sim6.8-6.9$~keV) and He$\beta$ triplet from Fe XXV (around $\sim7.75$~keV), several weak emission lines from highly ionized Fe ions, one weak emission feature from Ni XXVII, even weaker fluorescence lines of neutral Fe I and He$\alpha$ triplets from Cr XXIII and Mn XXIV \cite{hcl16}. 


Advance in technology certainly provides a better view of the hot plasmas in the Unvierse \cite{kaa17}. Benchmarking the \textit{Hitomi}/SXS spectrum of the Perseus galaxy cluster with collisional ionized plasma models widely used in the community reveals that accurate plasma models and the underlying atomic data are crucial to determine the physical properties (temperature, emission measure, and abundance, etc.) of the intracluster medium of Perseus. That is to say, improvement of plasma models and atomic data can exert an impact on our understanding of the X-ray Universe. In the following, we focus on two examples, elemental abundance of intracluster media and density of photoionized outflows, in the context of outdated and updated plasma models, as well as current and future generations of spectrometers.

\section{Elemental abundance of intracluster media}
A galaxy cluster typically consists of hundreds of member galaxies\footnote{A smaller aggregate of galaxies is referred to as a galaxy group.}, with a total mass of $10^{14-15}~{\rm M_{\odot}}$. A large fraction ($\sim15-20$~\%) of the mass is in the form of hot plasma with $kT\sim10^{7-8}~{\rm K}$, namely the intracluster medium (ICM), while the visible member galaxies contribute less than $\sim5$~\% \citep{sch08,and10,bud14}. The rest of the mass is in the form of dark matter.

A typical X-ray spectrum of the ICM in a relaxed cluster contains emission lines from various elements (e.g., N, O, Mg, Si, S, Fe, and Ni). These elements are originally synthesized in various types of stars within the cluster:
\begin{itemize}
    \item Asymptotic giant branch (AGB) stars are low- and intermediate mass stars with $M_{*}\in(0.9,~7)~{\rm M_{\odot}}$ undergoing the last nuclear burning phase. In this phase, the stellar envelope is enriched with products of hydrogen and helium burning, as well as heavy elements produced by the slow neutron capture process \cite{kar10}. These elements are eventually expelled into the interstellar medium (ISM) via a slow stellar wind.

    \item Massive stars in the mass range of $M_{*}\in(11,~140)~{\rm M_{\odot}}$ undergo Fe core collapse at the end of their stellar evolution \citep{nom13}. If the entire star does not collapse into a black hole with no mass ejection, it explodes as core-collapse supernovae (SNcc) and release elements into the ISM. 

    

    \item Degenerate stars in a binary system also contribute to the chemical enrichment of the ISM, mainly via Type Ia supernovae (SNIa) \citep{nom13}. 
\end{itemize}

\begin{figure}
\centering 
\includegraphics[width=.7\textwidth,trim=0 0 0 0, clip]{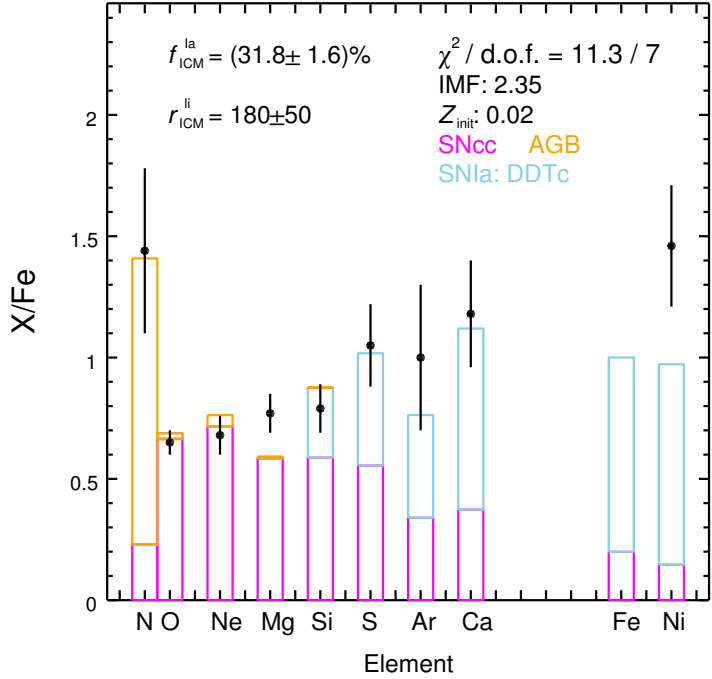}
\caption{\label{fig:ngc5044} The chemical enrichment of the hot plasma in NGC 5044 observed with \textit{XMM-Newton}. The abundance ratios are measured with the \textit{cie} model in SPEX v3.04. Under a standard initial mass function (IMF) with a slope of $-2.35$ \cite{sal55} and an initial metallicity ($Z_{\rm init}=0.02$, i.e. solar composition) of progenitors, a suitable mixture of different enrichment channels are required to fit the observed abundance ratios. Nitrogen is mainly enriched via the stellar winds of AGB stars. Core-collapse supernovae (SNcc) are the main elemental factory of the so-called $\alpha$-elements (e.g., O, Ne, and Mg). Type Ia supernovae (SNIa) dominate the enrichment of the Fe-peak elements (e.g., Fe and Ni). This figure, originally shown in \cite{mao19}, is reproduced with permission. }
\end{figure}

Throughout the evolution of the galaxy cluster, different transportation mechanisms working on different size-scales and time-scales redistribute these elements from the small-scale ISM into the large-scale ICM \cite{sch08}. Different enrichment channels (AGBs, SNcc, and SNIa) leave specific abundance patterns so that the observed (time-integrated) ICM elemental abundance can be used to infer the origin of these elements, as shown in Figure~\ref{fig:ngc5044}. 

Abundances are measured by fitting the observed spectrum with collision ionized equilibrium (CIE) plasma models. Evolved from early works in the 1970s, two CIE models are widely used nowadays for X-ray spectral analysis of the ICM: APEC \cite{smi01} and \textit{cie} in SPEX\footnote{SPEX is a spectral analysis package, which contains plasma models like \textit{cie} (for collisional ionized plasmas) and \textit{pion} (for photoionized plasmas). APEC is a standalone plasma model that can be used by spectral analysis packages like XSPEC \cite{arn96}.} \cite{kaa96}. 

Nevertheless, those pre-launch versions of APEC (v3.02) and SPEX (v3.00) do not provide a good fit to the newly obtained Perseus spectrum by \textit{Hitomi} in 2016. The mismatch triggers the implementation of some necessary updates, including extending incomplete atomic database, replacing less accurate atomic data, and improving the level population calculation \cite{hcl18}. Updated plasma models not only better fit the observed \textit{Hitomi} spectra of Perseus \cite{hcl18}, but also lead to a revised interpretation of spectra observed with CCD type of spectrometers \footnote{Generally speaking, the spectral resolution of microcalorimeters ($<5$~eV) is at least an order of magnitude better than that of CCD spectrometers ($50-200$~eV).} like the European Photo Imaging Camera (EPIC) aboard \textit{XMM-Newton} \cite{mer18b}. 

Prior to the launch of \textit{Hitomi}, 44 nearby relaxed galaxy groups and clusters were observed with \textit{XMM-Newton} to study their chemical composition and origin (referred to as the CHEERS sample hereafter) \cite{dpl17}. The \textit{cie} model in SPEX v2.06 was used to fit the EPIC spectra of the CHEERS sample \cite{mer16a}. Abundance ratios with respect to Fe were obtained and used to infer the origin of the elements \cite{mer16b}. Later, the updated \textit{cie} model in SPEX v3.04 was available, \cite{mer18b} re-estimate the abundance ratios of the CHEERS sample. Table~\ref{tbl:cheers} presents the comparison between the two abundance measurements. While the rest of the abundance ratios are consistent within uncertainties between the two versions of plasma code, the Cr/Fe and Ni/Fe ratios are closer to unity (i.e. protosolar abundance ratio) in the new version. This is mainly due to the fact that the new plasma model includes hundreds of thousands of weak lines. 

\begin{table}[htbp]
\centering
\caption{\label{tbl:cheers} Averaged abundance ratios with respect to Fe for the CHEERS sample. The original measurements were obtained with SPEX v2.06 as reported in \cite{mer16a}. The new version of SPEX (v3.04) was used by \cite{mer18b} to re-estimate the abundance ratios. }
\smallskip
\begin{tabular}{|c|c|c||c|c|c|}
\hline
Ratio & SPEX v2.06 & SPEX v3.04 & Ratio & SPEX v2.06 & SPEX v3.04 \\
\hline
O/Fe & $0.82\pm0.15$ & $0.82\pm0.18$ & Ne/Fe & $0.72\pm0.16$ & $0.72\pm0.13$ \\
Mg/Fe & $0.74\pm0.17$ & $0.94\pm0.07$ & Si/Fe & $0.87\pm0.06$ & $0.95\pm0.06$ \\
S/Fe & $0.98\pm0.10$ & $1.00\pm0.02$ & Ar/Fe & $0.88\pm0.15$ & $0.98\pm0.08$ \\
Ca/Fe & $1.22\pm0.10$ & $1.27\pm0.10$ & Cr/Fe & $1.56\pm0.19$ & $0.99\pm0.19$ \\
Mn/Fe & $1.70\pm0.22$ & $1.56\pm0.77$ & Ni/Fe & $1.93\pm0.40$ & $0.96\pm0.38$ \\
\hline
\end{tabular}
\end{table}


After the revision, the solar abundance ratios in the CHEERS sample (observed with \textit{XMM-Newton}/EPIC) agree with measurements from the \textit{Hitomi}/SXS spectrum of Perseus \cite{hcl17,sim19}, indicating a common solar composition of the hot plasmas in the core regions of nearby relaxed groups and clusters of galaxies \cite{mer18b}. In addition, setting aside the uncertainties in the chemical enrichment models and nucleosynthesis yields \cite{sim19}, the new abundance ratios suggest that near-Chandrasekhar-mass type Ia supernovae contribute significantly to the observed solar abundance ratios \cite{mer18c}. 


\section{Density of photoionized outflows}
At the center of almost every galaxy but the smallest lies a supermassive black hole (SMBH) with $M_{\rm BH}> 10^5~{\rm M_{\odot}}$ \cite{net15}. The growth of the SMBH is realized via accretion of matter. When the accretion rate is above a certain limit ($\gtrsim10^{-5}$ Eddington ratio), the central region of the galaxy is called an active galactic nucleus (AGN)  \cite{net15}.

While AGNs are powered by accretion (or inflows), outflows transporting matter and energy away from the nucleus to the host galaxy and beyond are also observed \cite{cre03}. More importantly, the discovery of the tight correlation between the mass of SMBH and the observables (velocity dispersion and K-band luminosity) of the bulge component of the host galaxy \cite{kor13} suggest that AGN-driven outflows might be the key to understand the evolution of SMBH and host galaxy \cite{mag98,sil98,lak19,gas19}. 

Various forms of AGN outflows have been observed in the past two decades \cite{fab12,har18}. In the X-ray band, outflows have been observed in many active galaxies \cite{tom13,lah14}. Nevertheless, it is rather uncertain whether these X-ray outflows are powerful enough to play an important role in the interplay between the SMBH and its host galaxy. 

The kinetic power carried by X-ray outflows can be estimated if the distance of the outflow with respect to the SMBH is known. Direct measurement of the outflow distance with respect to the black hole is not available. The size scale of SMBH is too small and the host galaxies are too far away so that it cannot be resolved via direct imaging. 


Density sensitive metastable absorption lines can be used to measure the density of the outflows. In a low density (photoionized) plasma, the level population of each ion is concentrated in the ground level, so that only absorption lines from the ground levels can be observed. As the plasma density increases, the collision between the free electrons and ions can populate the metastable levels, which leads to the presence of metastable absorption line(s) in the spectrum. As the total level population is conserved for each ion, the strong the metastable absorption line(s), the weaker the ground absorption line. 

For observers, a spectrometer with a sufficient spectral resolution and an adequate photon collecting area is essential. The former ensures that the metastable level absorption lines are clearly separated from the ground level absorption lines, ideally with both line profiles resolved. The latter ensures a high signal to noise ratio of the absorption features with reasonable exposure (up to a few hundreds of kilo-seconds). The current generation of grating spectrometers \cite{bri00,dhe01,can05} are not very effective for the density measurement with metastable level absorption lines. While a tight constrain on the density of a disk wind in a stellar mass black hole GRO J1655-40 \cite{mil08}, only upper or lower limits are obtained for densities of AGN outflows\cite{kaa04,kin12,mao17b}.

\begin{figure}
\centering 
\includegraphics[width=.9\textwidth,trim=0 0 0 0,clip]{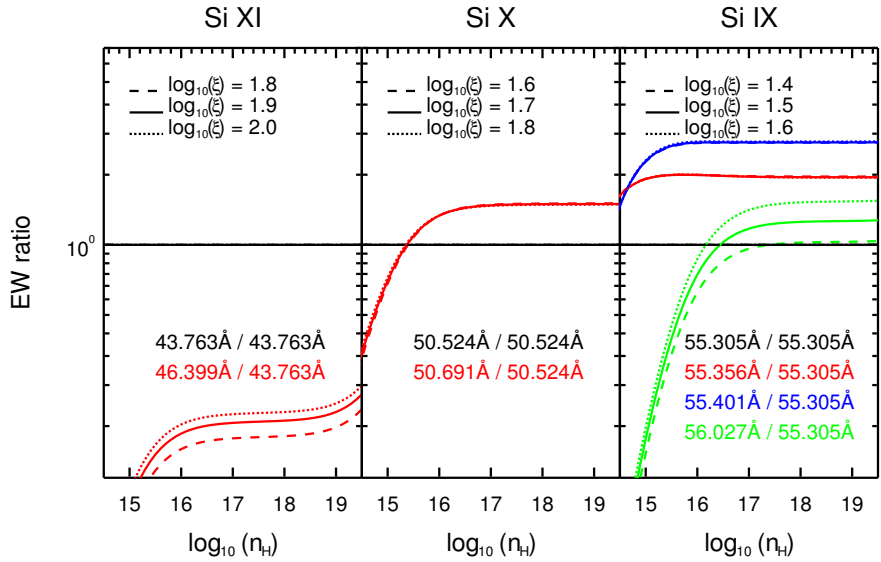}
\caption{\label{fig:msal} The equivalent width (EW) ratio of the metastable level absorption line to the ground level absorption line as a function of plasma density ($n_{\rm H}$) for (Be-like) Si XI, (B-like) Si X, and (C-like) Si IX \cite{mao17b}. Different line styles indicate different ionization parameter ($\xi$) of the photoionized plasma.}
\end{figure}

On the other hand, the density measurement relies on a photoionized plasma model built on an extensive atomic database. Inaccurate metastable to ground level population ratio can lead to biased estimations of density, thus, distance and kinetic power of AGN outflows. Prior to the release of the \textit{pion} model in SPEX v3.04, the collisional ionized equilibrium (CIE) model in the CHIANTI code was widely used to estimate the metastable to ground level population ratio. An assumption of the temperature of the CIE plasma is required to mimic the ionization balance of the photoionized outflow. Moreover, metastable levels and lines are not fully resolved in those early versions of CHIANTI. For instance, in the study of the ionized outflow in NGC 4051, \cite{kin12} used CHIANTI v6 \cite{der09} to estimate the ratio of the equivalent width of the metastable absorption line at 12.38 \AA\ to the ground absorption line at 12.29 \AA\ of C-like Fe XXI. In fact, for C-like ions (including Fe XXI), three metastable absorption lines can be observed (provided with adequate spectral resolution) from three metastable levels ($2s^2 2p^2,~^3P_1$, $^3P_2$, and $^3D_2$). For Fe XXI, the most prominent metastable absorption lines are 12.433 \AA\ (with the oscillator strength $f=0.43$), 12.331 \AA\ ($f=0.59$), and 12.411 \AA\ ($f=0.94$) \cite{mao17b}. Moreover, the population of one of the metastable levels is sensitive to both the density and ionization parameter (or temperature) of the plasmas (Figure~\ref{fig:msal}). A re-analysis of the ionized outflow in NGC 4051 is under review.

\section{Future perspectives}

The high-resolution X-ray observations of Perseus galaxy cluster with \textit{Hitomi}/SXS certainly helped to identify issues with pre-launch plasma models, as well as the accuracy of the underlying atomic data. Note that the benchmark was limited to spectral features above 1.8~keV, because a 262-micron Be-like filter was used to block soft X-ray photons during the performance validating phase. Unfortunately, \textit{Hitomi} spun out of control less than two months since its launch date. Several \textit{Hitomi}-level missions (\textit{\href{https://heasarc.gsfc.nasa.gov/docs/xrism/}{XRISM}}, \textit{Athena} \cite{nan13}, \textit{Arcus} \cite{smi16}, and \textit{\href{http://heat.tsinghua.edu.cn/~hubs/en/index.html}{HUBS}}, etc.) have been proposed and some funded. 

\begin{figure}
\centering 
\includegraphics[width=.7\textwidth,trim=40 20 40 20,clip]{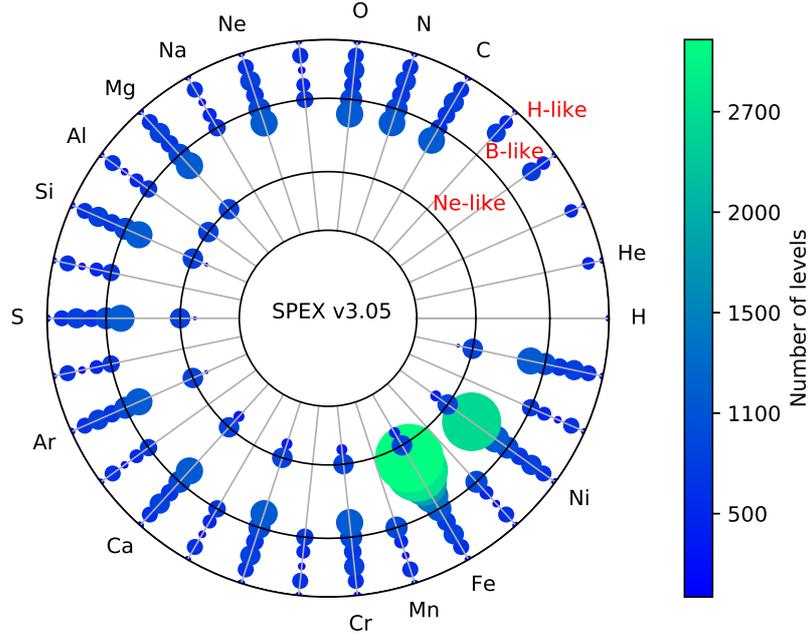}
\caption{\label{fig:spex_v305} The number of energy levels for each ion (color coded) included in the latest version of the SPEX code (v3.05). The (in)completeness can also be inferred from the size of the filled circle. Many level-resolved rates and cross-sections are available for H-like (outer most ring) to B-like ions with atomic number with $Z\le30$. Atomic data for C-like, Ne-like and Na-like ions are limited to cosmic abundance elements (e.g., C, N, O, Si, Fe, and Ni). Data for N-like to F-like data are lacking, except for Fe and Ni ions. }
\end{figure}

Plasma models need to be prepared for the high-quality X-ray spectra in the near future. Some efforts have been initiated in the community. The first one is to recognize the incompleteness of the current atomic databases. Level-resolved rates and cross-sections spanning several orders of magnitude in the parameter space (e.g. temperature and density) are the building blocks of the plasma model. Therefore, the number of levels for each ion included in a plasma model indicates the (in)completeness of atomic data. An impression of the number of energy levels for each ion included in the atomic database of SPEX v3.05 is illustrated in Figure~\ref{fig:spex_v305}. Generally speaking, elements with a even atomic number ($Z$) have more data as they are more abundant in astrophysical environments. Compared to H-like to B-like ions, atomic data of ions with six or more electrons are lacking, except for Fe and Ni ions. 

A similar level of incompleteness applies to the atomic database of other codes like APEC and CHIANTI. Furthermore, detailed rates, cross-sections and sometimes energy levels can differ significantly among these atomic databases. Without being complete, we list a few sources of discrepancy. First, the existence of outdated atomic data while recent and accurate atomic data are available. Second, some atomic data are only accurate in a limited range of parameter space but extrapolation can sometimes lead to undesired behavior. Third, several theoretical calculations might be available, yet do not agree with each other. Due to the lack of benchmarking with lab measurements, different plasma codes and models might take different calculations. Therefore, comparisons among different plasma models and codes have been made \cite{meh16,hcl18} to identify the discrepancy, especially those related to important transitions.  

To advance our knowledge of the Universe, better plasma models and atomic data are certainly required. Odd-$Z$ elements are important, for instance, their abundance are more sensitive to the initial metallicity of the progenitors of AGBs and SNe \cite{mao19}, which helps us to better understand the chemical enrichment of the Universe. The ground and metastable absorption lines of various ions are essential to probe AGN outflows in a wide range of density and distance \cite{mao17b}. The accuracy of plasma models correspond to spectral features below 1.8~keV (not observed with \textit{Hitomi}/SXS) needs to be investigated. The Fe abundance of low-temperature (and low-mass) elliptical galaxies and groups of galaxies is measured from the Fe-L band emission features below 1.8 keV. Two different trends of iron abundance with an increasing temperature/mass are reported using different (versions of) plasma models \cite{yat17,mer18a}, which requires re-investigation with the future plasma codes. The above examples are just a taste of the need of plasma models and atomic data. 

Once we know our need for atomic data, theoreticians can perform new calculations to complete the atomic database or improve the accuracy of the existing data \cite{mao17a,gu19}, new lab measurements can be triggered to benchmark theoretical calculations \cite{sha19}. Subsequently, modellers need to take advantage of more accurate atomic data and validate the implementation (e.g., parameterization, interpolation, and extrapolation) in the plasma models. The remaining discrepancy among different plasma models and codes can be used as a measure of systematic uncertainty. 



\acknowledgments
J.M. is supported by STFC (UK) through the University of Strathclyde UK APAP network grant ST/R000743/1.
L.G. is supported by the RIKEN Special Postdoctoral Researcher Program. 
F. M. is supported by the Lend\"{u}let LP2016-11 grant awarded by the Hungarian Academy of Sciences. 
M.M. is supported by NWO through the Innovational Research Incentives Scheme Vidi grant 639.042.525.
SRON is supported financially by NWO, the Netherlands Organization for Scientific Research.



\end{document}